\documentclass[fleqn,usenatbib]{mnras}
\usepackage[T1]{fontenc}
\usepackage{color}
\usepackage{graphicx}
\usepackage{bm}
\usepackage{amsmath,amsfonts,amssymb}

\begin{document}

\title{Physics of star formation history and 
the luminosity function of galaxies therefrom} 

\author[M. Fukugita and M. Kawasaki]{Masataka Fukugita$^{1}$ and Masahiro Kawasaki$^2$
\\
$^1$Kavli Institute for the Physics and ~Mathematics of the Universe, University of Tokyo, Kashiwa, 277-8583, Japan\\
$^2$Institute for Cosmic Ray Research, University of Tokyo, Kashiwa, 277-8582, Japan}


\maketitle

\begin{abstract}
We show that the star formation history, the reionization history and
the present luminosity function of galaxies are reproduced
in a simple gravitational collapse model within the $\Lambda$CDM regime to almost a quantitative accuracy, when the physical conditions,
the Jeans criterion and the cooling process, are taken into account.
Taking a reasonable set of the model parameters, the reionisation takes place sharply at around redshift $1+z\simeq 7.5$, and the resulting luminosity function turns off at $L\simeq 10^{10.7}L_\odot$, showing the consistency between the star formation history and the reionisation of the Universe.
The model gives the total amount of stars $\Omega_\mathrm{star}=0.004$ in units of the critical density compared to the observation $0.0044$ with the recycling factor $1.6$ included.
In order to account for the observed star formation rate and the present luminosity function, the star formation efficiency is not halo mass independent but becomes maximum at the halo mass $\simeq 10^{12}M_\odot$ and is suppressed
for both smaller and larger mass haloes.

\end{abstract}



\section{Introduction}

$\Lambda$CDM model gives the basis of cosmology and the large-scale
structure of the universe through gravitational instability of dark
matter: the cosmological parameters are precisely determined from the
cosmic microwave background (CMB)~\citep{Planck:2018vyg}.
Density fluctuations of dark matter grow and
collapse into haloes in a hierarchical manner with small
haloes produced first and larger haloes formed
subsequently merging small halos, as clearly seen in the
semi-analytical approach~\citep{Press:1973iz,Sheth:1999mn} and also $N$-body
simulations.

Baryons also collapsed in dark matter haloes when gravity overcomes
baryonic pressure; they are expected then to cool and form stars.  This
basic idea for the star formation and the galaxy formation has been developed
for years, 
in e.g.~\citet{1977ApJ...211..638S,1977MNRAS.179..541R,Blumenthal:1984bp}.
Many papers have appeared using 
semi-analytic approach 
(e.g.~\citet{White:1991mr,Kauffmann:1993gv,Cole:1994ab}).
Among them
\citet{Fukugita:1993ds,1994ApJ...420..484T,2003MNRAS.343L..25F}
investigated reionization of the universe by OB stars.
Hydrodynamical
simulations~\citep{Benson:1999mv,Yoshida:2002gm,Helly:2002ph} 
are employed for investigating the galaxy formation and  
the large scale structure together with gas dynamics
under less assumptions and idealization.  The semi-analytic
approach, however, has advantage to make 
underlying physics clear, such as the importance of physical processes
involving electrons, baryons and photons that are crucial in
formation of stars and galaxies. In large dimensional parameter space
large-scale cosmological simulations require a huge computational cost
for varying input parameters. 
We remark that there are several pieces of the work in this direction, but with different strategies and foci~\citep{Trenti:2010sz,Dekel:2013uaa,Tacchella:2012ih,Mason:2015cna,2017MNRAS.464.1365M,2017MNRAS.472.1576F}.

In this paper we study the star formation history, as depicted in the
`Madau plot'~\citep{Madau:1997pg} and the luminosity
function of galaxies using the simple gravitational collapse model
in a semi-analytic treatment.
The mass function of dark matter haloes is computed using the 
Press-Schecheter approach~\citep{Press:1973iz}. We apply the cooling and
the Jeans conditions to the haloes to obtain star formation:
the galaxy haloes are identified when the
conditions are satisfied that produce stars.
We study the halo mass dependence of the star formation efficiency and
calculate evolution of the star formation rate, the reionization history
and the luminosity function.  We shall observe that the mass dependence
 is crucial in star formation efficiency
$f_\text{eff}$~ for successfully reproducing both observed
star formation history and luminosity function.

We note that each individual process has already been studied well
in the literature
to now. What we are going to do is \lq engineering', putting together the
processes and to find if this would produce the observation correctly, and
what conditions are needed. It is important that we can study the
dependence on the input parameters.
Compared with similar studies~\citep{Trenti:2010sz,Dekel:2013uaa,Tacchella:2012ih,Mason:2015cna,2017MNRAS.464.1365M,2017MNRAS.472.1576F} our focus is on the star formation history and the luminosity function along with the ionisation history.
In our calculation the luminosity-mass relations of main-sequence and red giant stars are used as input.

In Sec.~\ref{sec:simple-collapse} 
the formulae are assembled to calculate the star formation rate and
the luminosity function.
In Sec.~\ref{sec:result} 
we present the star formation history, the reionization history and
the luminosity function of galaxies. 
Sec.~\ref{sec:conclusion} presents the conclusion we obtained. 
In this paper we take $\Lambda$CDM model with the cosmological parameters 
\begin{equation}
    \begin{array}{lll}
        h = 0.673   \qquad &  \Omega_m h^2 =0.143 \qquad & \Omega_b h^2 = 0.0224 \\[0.6em]
        n_s = 0.965 &  \sigma_8 = 0.811 & Y_p = 0.247
    \end{array}
\end{equation}
for the Hubble constant, the total matter density, the baryon density,
the spectral index, the fluctuation amplitude at $8h^{-1}$~Mpc, and
the helium abundance, as taken from Planck 2018~\citep{Planck:2018vyg}.

\section{Simple gravitational collpse model}
\label{sec:simple-collapse}

\subsection{Basic formulae used}
The halo mass function may be calculated with the Press-Schechter formula~\citep{Press:1973iz}. 
Here we adopt a more accurete fitting formula by Sheth and Tormen~\citep{Sheth:1999mn}.  
The halo comoving number density $N(M,z)$ with mass $M\sim M+dM$ is written as
\begin{align}
n(M,z) &= A
    \sqrt{\frac{2}{\pi}}\frac{\rho_{m0}}{M}
    \left(\frac{a\delta_{c}^2}{\sigma(M)^{2}}\right)^{1/2}
    \left(1+\left(\frac{a\delta_c}{\sigma(M)^2}\right)^{-p}\right)\nonumber\\
    & \left|\frac{\partial\ln \sigma(M)}{\partial M}\right|
    \exp\left(-\frac{a\delta_{c}^{2}}{2\sigma(M)^{2}}\right),
\end{align}
where $\rho_{m0}$ is the present matter density, $\delta_c = 1.686$, $a=0.707$, $p=0.3$, $A=0.322$ and $\sigma(M)$ is the coarse-grained variance.
In working details we follow the formalism of \citet{Bardeen:1985tr}.


Baryons collapse in dark matter haloes, if the gravity overcomes the
baryon pressure, namely if the halo mass is larger than
the Jeans mass $M_\text{J}$,

\begin{equation}
   \label{eq:Jeans}
   M_\text{J} = 1.4\times 10^4 M_\odot \, (\Omega_m h^2)^{-1/2}
   \left(\frac{T_e}{\mu T}\right)^{3/2},	
\end{equation}
where $T$ is the cosmic temperature, $T_e$ the electron temperature,
and $\mu$ the mean molecular weight.

Star formation takes place when baryons 
lose their energy through atomic and molecular cooling.
The atomic cooling of hydrogen and helium becomes effective when
the virial temperature $T_\text{vir}$ is larger than about $10^4$~K, where
%
\begin{equation}
    \label{eq:virial_temp}
    T_\text{vir} = 0.01\,\text{K} \left(\frac{M}{M_\odot}\right)^{2/3}
    \mu (1+z) (\Omega_m h^2)^{1/3}\left(\frac{\rho/\bar{\rho}}{178}\right)^{1/3},
\end{equation}
with $\rho\, (\bar{\rho})$ the halo (background) density. 
Eq.~\eqref{eq:virial_temp} indicates that the atomic cooling works if 
\begin{equation}
    \label{eq:atomic_cooling}
    M > M_\text{ac} = 2.1\times 10^9 \, M_\odot
     (\Omega_m h^2)^{-1/2}(1+z)^{-3/2}.
\end{equation}
Here we take $\mu=1.22$ and $\rho/\bar{\rho}=178$.
The molecular cooling condition~\citep{Blumenthal:1984bp}
is 
\begin{align}
    \label{eq:mol_cooling}
    M > M_\text{mc} = & 2.1\times 10^8 \, M_\odot\,
    (\Omega_m h^2)^{-0.917}\nonumber \\
    & \times \left(\frac{Y_e}{10^{-4}}\right)^{-0.625}
    \left(\frac{\Omega_b}{\Omega_m}\right)^{-2.04}
    (1+z)^{-2.75},
\end{align}
where $Y_e$ is the fraction of free electrons.

For large haloes with virial temperature $T_\text{vir} \gtrsim 10^6$~K,
the cooling rate is also controlled by metal cooling or Compton cooling.
The metal cooling rate~\citep{Dalgarno:1972ak} is given by
\begin{equation}
    \Lambda_\text{metal} \simeq 1.45\times 10^{-7}
    \left(\frac{Z}{Z_\odot}\right)~ 
    \text{K\,cm}^3\,\text{s}^{-1}
    \quad  (T \gtrsim 10^{6}\,\text{K}),
\end{equation}
with $Z(Z_\odot)$ the (solar) metallicity.
For efficient cooling in haloes,
$T_\text{vir}/(n_B \Lambda_\text{metal}) \lesssim H^{-1}$,
which leads to 
\begin{equation}
    M < M_\text{metal} = 3.0\times 10^{14}\, M_\odot\,
    (\Omega_b h^2)^{3/2}(\Omega_m h^2)^{5/4}
    \left(\frac{Z}{0.01}\right)^{3/2}.
\end{equation}
On the other hand, Compton cooling is efficient if 
\begin{equation}
    1+z \gtrsim 1+z_\text{comp}=5.3\,\left(\frac{\Omega_m h^2}{0.25}\right)^{1/5}.
\end{equation}
Thus, we assume that star formation starts if the halo mass $M$ satisfies
\begin{align}
	 M_\text{min}
	 < M < M_\text{max},
\end{align}
where 
\begin{align}
    M_\text{min} 
    & = \max\left[M_\text{J}, \,\min(M_\text{ac},M_\text{mc})\right] 
    \\[0.5em]
    \label{eq:Mmax}
    M_\text{max}
    &= \begin{cases}
        M_\text{metal} & (z<z_\text{comp})  \\
        \infty         & (z>z_\text{comp})~.
        \end{cases}
\end{align}

In calculating the star formation we assume that the formation rate is proportional to the derivative of the halo mass function, $\partial n(M,t)/\partial t$ which is considered as the formation rate of halos with mass $M$.
In the Press-Schechter theory, halos are formed through accretion and merger.
Thus, $\partial n(M,t)/\partial t$ is proportional to the rate of accretion
into halos when the accretion is the dominant process.\footnote{
If we consider accretion only, the halo mass function evolves as 
$N(M,t) = N(M-\dot{M}dt, t-dt)$, which leads to $\partial N/\partial t = \dot{M}(-\partial N/\partial M)$.
Here $\dot{M}$ is the accretion rate of dark matter into a halo with mass $M$.
}
In this paper, we neglect the effect of the merger.

We assume that stars are produced in two ways:
the burst star formation takes place when the haloes formed with the rate
\begin{equation}
     \psi_\text{b}(z) = \int_{M_\text{min}}^{M_\text{max}} dM \, 
     F_\text{b}f_\text{eff}(M)\, 
     \frac{\Omega_b}{\Omega_m} M\frac{\partial n(M,z)}{\partial t},     
\end{equation}
where $F_\text{b}$ is the maximum fraction of collapsed baryons that form stars
in a burst and $f_\text{eff}\,F_\text{b}$ is the efficiency of star formation
which may depend on the halo mass $M$.
With the baryons survived the burst 
stars formed continuously in haloes.
We assume that stars form at a constant rate
per baryon mass for the duration $\tau_\text{c}$.
The star formation rate is then written 
\begin{equation}
    \label{eq:cont_sfr}
    \psi_\text{c}(z) = 
    \int^{t(z)}_{t(z)-\tau_\text{c}} dt' 
    \int_{M_\text{min}}^{M_\text{max}} dM \,
    \frac{\Omega_b}{\Omega_m}
    \dot{F}_\text{c}f_\text{eff}(M)
    \frac{\partial n(M,z(t'))}{\partial t'},
\end{equation}
where $\dot{F}_\text{c}$ is the maximum star formation rate per
baryonic mass in a halo.  The observation suggests that star formation
is less efficient for low mass galaxies. Massive galaxies with
$M\gtrsim 10^{13}M_\odot$ are rare too. In order to take account of these
suppressions,  we introduce the star forming efficiency $f_\text{eff}$,
assuming that it
 depends on the halo mass, as
\begin{equation}
    \label{eq:sf_eff_b}
    f_\text{eff}  = 
    \begin{cases}
        (M/M_*)^{\alpha_\text{L}} & \quad \text{for} \quad M \le  M_*\\
        (M/M_*)^{-\alpha_\text{H}} & \quad \text{for} \quad M > M_* ,
    \end{cases}
\end{equation}
where we take $M_{*} = 10^{12}M_\odot$;
we also take a default $\alpha_\text{L}\sim 0.7$
and $\alpha_\text{H}\sim 1$, as discuss below. 
The total star fomation rate $\psi_\text{sfr}$ is given by the sum
\begin{equation}
    \psi_\text{sfr} = \psi_\text{b}+\psi_\text{c}.
\end{equation}

\subsection{UV photons}


The UV photons emitted from stars with mass larger than $\sim
10M_\odot$ reionize the universe and heat the intergalatic medium,
which then affects star formation through the modification of the Jeans mass.
We adopt the~\cite{Chabrier:2003ki} initial mass function (IMF) 
\begin{align}
    \log_{10}\phi(m_s) =
    \begin{cases}
        A- 0.912\log_{10}(m_s/M_\odot) \\
         \quad - 0.456\,[\log_{10}(m_s/M_\odot)]^2 & (m_s < 1M_\odot),\\[0.5em]
        A- 1.3\log_{10}(m_s/M_\odot) & (m_s \ge 1M_\odot),
    \end{cases}   
\end{align}
where $m_s$ is the mass of stars.
$A$ is the normalisation so that $\int dm_s \phi(m_s) = 1$.
The stellar temperature-mass relation for Population II ($Z < 0.001$)
and I ($Z>0.01$) stars
is given by~\citep{Bond:1985pc}
\begin{equation}
    T_s(m_s) = 
    \begin{cases}
       6\times 10^4 \,\text{K}\, 
       \min\left[ \left(\frac{m_s}{100M_\odot}\right)^{0.3},1\right]
       &  ( Z < 0.001)\\[0.8em]
       4.3\times 10^4 \,\text{K}\, 
       \min\left[ \left(\frac{m_s}{100M_\odot}\right)^{0.3},1\right]
       &  ( Z > 0.01),
    \end{cases}
\end{equation}
respectively.  For $0.001 < Z < 0.01$ we consider a mixture of two
populations with the Population~I fraction $(Z-0.001)/0.009$.


Lifetime of stars with $m_s \gtrsim 10M_\odot$ that emits ionising UV
is shorter than $10^8$~yr, shorter than the Hubble time at $z < 20$.
Therefore, we may assume that UV photons are emitted instantaneously
upon the formation of stars. 
UV photons with energy $\varepsilon_\gamma$ are produced, as
\begin{align}
    \label{eq:uv_flux}
    \frac{d n_\gamma (\varepsilon_\gamma,z)}{dt} & = f_\text{esc}
    \int dm_s \frac{B(\varepsilon_\gamma,T_s)}{\varepsilon_\gamma}
    \epsilon_s \phi(m_s)\frac{\Omega_b}{\Omega_m} \psi_\text{sfr}(z),
\end{align}
where $B(\varepsilon_\gamma,T_s)$ is the blackbody spectrum normalized
as $\int d\varepsilon_\gamma B(\varepsilon_\gamma) =1$, and
$\epsilon_s$ is the fraction of the radiation energy to the rest mass.

Here, we introduce the escape fraction $f_\text{esc}$ to account for
absorption of UV photons while escaping
the halo.
We follow the evolution of the UV photon spectrum $n_\gamma(\epsilon_\gamma)$ by numerically integrating the Boltzmann equation with source term given by Eq.~\eqref{eq:uv_flux}.
We calculate the cumulative effective ionizing flux, 
\begin{equation}
    \label{eq:ionization_flux}
    F_\text{ion}(z) = \int_{\varepsilon_\text{th}}^{\infty}
    d\varepsilon_\gamma\, 
    \frac{\sigma_{\rm ph}(\varepsilon_\gamma)}{\sigma_{\rm ph}(\varepsilon_\text{Th})}
    \int^{t}dt'
    \frac{d n_\gamma (\varepsilon_\gamma(1+z')/(1+z),z')}{dt'},
\end{equation}
where $\sigma_{\rm ph}$ is the photoionization cross section of hydrogen
and $\varepsilon_\text{th}=13.6$~eV.

We calculate the evolution of the etectron temperature taking into account photoionization, collisional ionization, Compton heating and cooling, also
including other minor processes as detailed in~\cite{Fukugita:1993ds}.
Furthermore, we solve the evolution equation for fractions of HI, HII, HeI, HeII, and HeIII.

We follow the evolution of metallicity.
Stars with mass $m_s > 4M_\odot$ produce heavy elements
at a fraction~\citep{1984ApJ...277..445C}
\begin{align}
    Z_\text{ej}  = 
    \begin{cases}
        0.5-(m_s/6.3M_\odot)^{-1} & 
          (15 < m_s/M_\odot < 100) \\[0.5em]
        0.1 & 
          (8 < m_s/M_\odot < 15) \\[0.5em]
        0.2 & 
          (4 < m_s/M_\odot < 8) .
    \end{cases}
\end{align}
The evolution of metallicity is 
\begin{equation}
    \frac{dZ}{dt} = \frac{1}{\rho_{b0}}
    \int dm_s \, Z_\text{ej}\phi(m_s) \psi_\text{sfr}(z).
\end{equation}

\subsection{Luminosity function}

For the luminosity function of galaxies
long-lived stars with mass $m_s \lesssim 10M_\odot$
are important.
We adopt the luminosity-mass relation of the main-sequence stars
in~\citet{2002AJ....124.2721R}, where the stellar mass is translated
to luminosity and represented
as absolute magnitude. 
We obtain the inverse relation, magnitude as a function of stellar mass, as 
\begin{align}
     M_V -M_{V\odot}& =  -13.463\,x + 0.823\, x^2 +5.920\,x^3 \nonumber \\
         & \quad - 0.570\, x^4 + 1.232\,x^5, 
\end{align}
where $x = \log_{10}(m_s/M_\odot)$, $M_V$
is the absolute
magnitude for the V-band luminosity, 
$L_\text{MS} = 10^{-0.4(M_V-M_{V\odot})}L_\odot$.
Here we consider only stars with mass less than $10 M_\odot$,
whose main sequence time is given by
\begin{equation}
    \log_{10}(t_\text{MS}/\text{yr}) = 10.03 -3.45\,x
     +0.786\,x^2 ~~ (m_s < 10M_\odot)
\end{equation}
For low mass stars with $m_s < 2M_\odot$ the time spent in the red giant
phase is significant compared to $t_\text{MS}$.
The red giant branch time $t_\text{RGB}$ ~\citep{Serenelli:2006gb}
may be fitted as
\begin{align}
    \log_{10}(t_\text{RGB}/\text{yr}) & = 9.514 -3.907\,x
     -19.50\,x^2 
     +43.60\,x^3  \nonumber \\
     & \quad\quad\quad (m_s < 2M_\odot),   
\end{align}
and luminosity of stars in the red giant branch is 
\begin{align}
    \log_{10}(L_\text{RGB}/L_\text{MS}) & =
    8.63 -18.4\, (m_s/M_\odot)  \nonumber \\
     & +14.2\, (m_s/M_\odot)^2 -3.51\, (m_s/M_\odot)^3.
\end{align}
The luminosity to mass ratio $L/M$ of the galaxy 
formed at $t_f$ is written 
\begin{align}
	\left(\frac{L}{M}\right)(t,t_f) & = \nonumber \\
        F_b \frac{\Omega_b}{\Omega_m} &
        \biggl[\int^{10M_\odot}dm_s\, 
        \theta[t_\text{MS}(m_s)-(t-t_f)] 
        L_\text{MS}(m_s)\phi(m_s) \nonumber \\[0.5em]
        + &\int^{2M_\odot} dm_s\,
        \theta[t_\text{MS}(m_s)+t_\text{RGB}(m_s)-(t-t_f)] \nonumber \\
        & \times \theta[(t-t_f)-t_\text{MS}(m_s)] 
        L_\text{RGB}(m_s)\phi(m_s) \biggr]
        \nonumber \\[0.5em]
	    +\dot{F}_\text{c} \frac{\Omega_b}{\Omega_m} &
        \int_{t_f}^{\min(t,t_f+\tau_\text{c})} dt' \nonumber \\
        & \biggl[
	    \int^{10M_\odot} dm_s\,
        \theta[t_\text{MS}(m_s) -(t-t')] L_\text{MS}(m_s)\phi(m_s)
        \nonumber\\[0.5em]
        &+\int^{2M_\odot} dm_s\,
        \theta[t_\text{MS}(m_s)+t_\text{RGB}(m_s)-(t-t')] \nonumber \\
        & \times \theta[(t-t')-t_\text{MS}(m_s)] 
        L_\text{RGB}(m_s)\phi(m_s) \biggr].
        \nonumber \\[0.5em]
\end{align}
The luminosity function is 
\begin{equation}
   \label{eq:luminosity_func}
	\Phi(L,z) = \int^{t(z)}dt' \int dM 
	\left(\frac{L}{M}\right)(t,t') 
	M \frac{\partial n(M,z(t'))}{\partial t'}.
\end{equation}
Note that $\left(\frac{L}{M}\right)(t,t_f)$ is a decreasing function of
$t$; massive stars ($m_s >2M_\odot$) with the main-sequence time
shorter than $t-t_\text{sf}$ ($t_\text{sf}$: star formation time)
do not contribute to luminosity.


\section{Results}
\label{sec:result}

Our calculation contains 7 model parameters
$(\alpha_\text{H}, \alpha_\text{L}, M_*, F_\text{b}, \dot{F}_\text{c}, \tau_\text{c}, f_\text{esc})$.
%
The stellar mass function of
galaxies~\citep{Bernardi:2013mqa,2017MNRAS.470..283W} turns off
at around $ M_*\sim (10^{11}-10^{12})M_\odot$. 
For low masses $M<M_*$
it is indicated that the mass function is suppressed by a factor
$M^{-0.4}-M^{-0.7}$ relative to the mass function of dark
matter~\citep{Bullock:2017xww}. 
For galaxies $\gtrsim 10^{11}M_\odot$, on the other hand,
the suppression of the stellar mass
function is given approximately   by 
$\alpha_\text{H} \sim 1.0$~\citep{Bullock:2017xww}
relative to dark matter.
Here we remark that, unlike the suppression factor in ~\cite{Bullock:2017xww}, our $\alpha_\text{H(L)}$ does not include suppression of star formation coming from the cooling and the Jeans conditions, which may lead to smaller $\alpha_\text{H(L)}$.

We take the fraction of the burst formation of stars to be $0.4$ to fit the star formation rate history better, 
which is somewhat larger than that implied by the age distribution of the stars in the solar neighborhood
~\citep{1988ApJ...334..436B}, which suggests
(burst)\,:\,(continuous)$= F_\text{b}:\dot{F}_\text{c}\tau_\text{c} \sim 0.3:0.7$.\footnote{Here we take stars with age less than $2.5$~Gyr
after the onset of collapse as formed in the burst.}
We take the duration of the star formation as $\tau_\text{c} \sim 10$~Gyr. 
The escape fraction $f_\text{esc}$ was measured by a number of
  authors, e.g.,
in~\citet{1995ApJ...454L..19L,Hurwitz:1997nm,Heckman:2001yx,Bouwens:2015vha} [see also, \cite{Robertson:2015uda,2016MNRAS.460..417S}], which vary but give
typically $f_\text{esc}\approx 0.03-0.1$.

From these observational implications and a comparison of our
calculation with the star formation and
reionization histories and with the galaxy luminosity function (which we
show below), we take
\begin{align}
    & \alpha_\text{L} = 0.67 , \quad \alpha_\text{H} = 0.5, 
    \quad M_* = 10^{12}M_\odot, \quad   F_\text{b} = 0.4\nonumber \\
    & \tau_\text{c}=3\times 10^{17}\,\text{s}, \quad
    \dot{F}_\text{c}=0.2\times 10^{-17}\,\text{s}^{-1}, \quad 
    f_\text{esc} = 0.06
\end{align}
as our optimal default set of the model parameters.

\subsection{Star formation rate}

\begin{figure}
   \centering
   \includegraphics[width=\columnwidth]{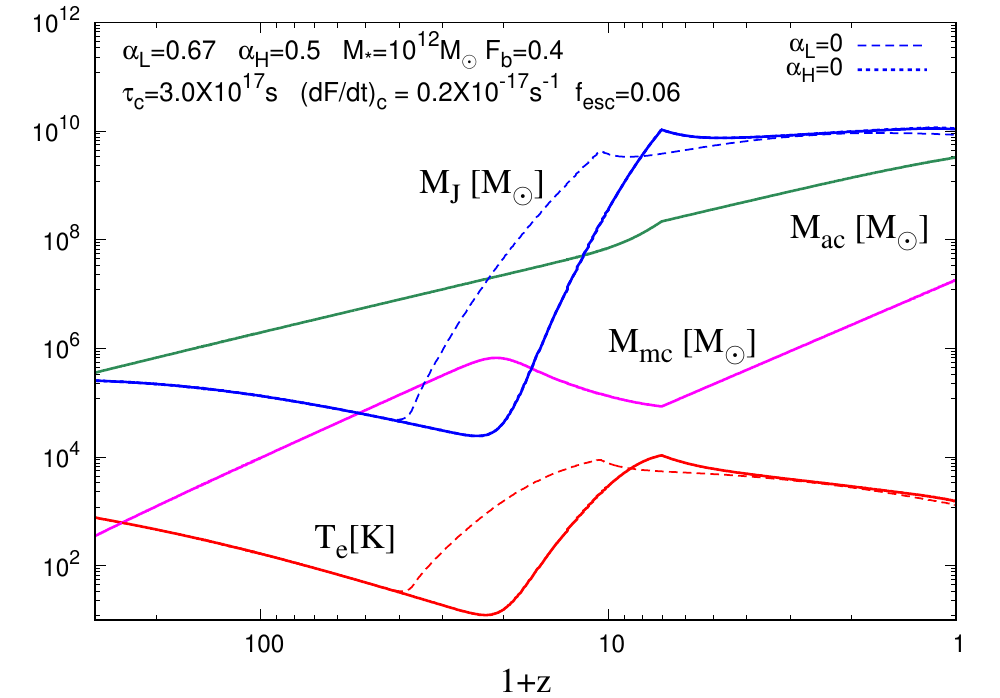}. 
   \caption{Evolution of the Jeans mass (blue), the lower bound from
     atomic cooling (green), the lower bound from atomic cooling (magenta),
     and of the electron temperature (red). 
     The Jeans mass and the electron temperature for $\alpha_\text{L}=0$ or
     $\alpha_\text{H}=0$ are also shown by dashed or dotted lines.
     The dotted lines may not be discernible as they almost
     degenerate with the curves for the default.}
   \label{fig:jeans_mass}   
\end{figure}

The minimum halo mass for
star formation is determined by the Jeans and cooling
criteria~\eqref{eq:Jeans}-\eqref{eq:mol_cooling},
which are depicted in Fig.~\ref{fig:jeans_mass}.
We see that the Jeans mass gives the lower bound for star formation
at $z \lesssim 10$, while the molecular cooling condition gives the
minimum mass for $10 \lesssim z \lesssim 50$. 
By $z \lesssim 6$ the Jeans mass rapidly increases to
$\simeq 10^{10}M_\odot$, which strongly suppresses the formation of star
forming galaxies. 
This rapid rise is caused by the increase of
the electron temperature due to heating with UV photons emitted from
early galaxies.

\begin{figure}
   \includegraphics[width=\columnwidth]{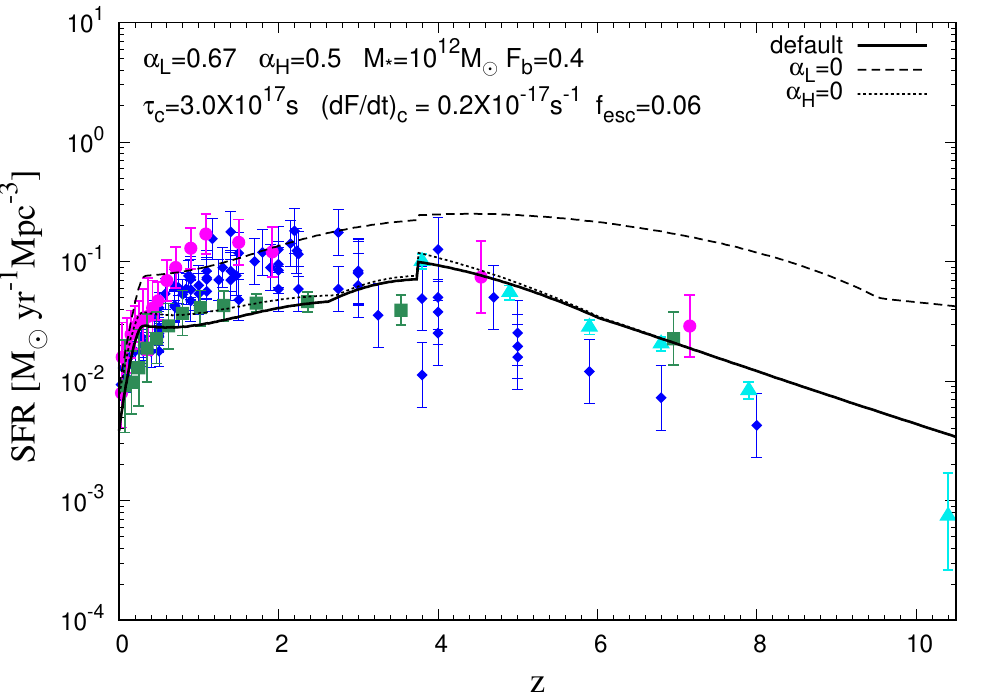}
   \caption{Cosmic star formation rate (solid line) compared with 
    the observation~\protect\citep{Behroozi:2012iw} 
    (blue points), \protect\citep{Bouwens:2014fua}(cyan points), \protect\citep{2019MNRAS.482.1557S}(magenta points) and \protect\citep{2018A&A...615A..27L}(green points). 
    The star formation rate  with the formation efficiency
    $(\alpha_\text{L}, \alpha_\text{H})=(0,1.4)$ and $(0.7,0)$
    are shown with dashed and dotted lines, respectively.}

   \label{fig:sfr}      
\end{figure}

We also include in this figure
the evolution of the Jeans mass and electron temperature
for $\alpha_\text{L}=0$ or $\alpha_\text{H}=0$ by dashed or dotted lines,
respectively.
The dotted lines for $\alpha_\text{H}=0$ are barely discernible  
from the default curve.
For $\alpha_\text{L}=0$, on the other hand, the electron temperature and
Jeans mass increase earlier,
since a large number of stars are produced in early-formed
low-mass haloes
without the suppression of the star formation efficiency
(see Fig.~\ref{fig:sfr}).
For $\alpha_\text{H}=0$ the star formation is enhanced close to the
present epoch, too large for $z\lesssim 0.3$.
This, however, hardly changes the electron temperature,
because the hydrogen is already fully ionized so that the energy
deposition by UV photons from stars is small.

The cosmic star formation rate for our default model parameters is
shown with solid line in Fig.~\ref{fig:sfr}.
This is compared with 
the recent compilation of
observations~\citep{Behroozi:2012iw}, which are shown in blue points in the
figure. 
Since our cosmic star formation rate is calculated without the luminosity cutoff, it overproduces the star formation rate compared to the observed data at high redshifts.
The estimated star formation rate history is consistent with
the observation within error bars.

The small kink of the curve seen at $z\sim 4$ is due to the Compton
cooling that decouples 
at this epoch, which changes $M_\text{max}$ rapidly
as seen in Eq.~\eqref{eq:Mmax}. 
This effect, however, is not important, because few large halos
formed at a high redshift. 
The star formation rate shows a rapid decrease around $z\sim 0.3$.
This comes from our choice of the star formation duration $\tau_\text{c}$,
or $t_0-\tau_c$
which corresponds to $z\sim 0.3$.
The star formation ceases in galaxies formed early, which accounts
for the rapid decrease of the star formation rate, consistent
with the decrease seen in what is observed.
In other words, the duration of the star formation
$\tau_\text{c}\simeq 10$~Gyr is required from comparison with the observation.

The efficiency for star formation that depends on the halo mass 
is represented by the parameters
$\alpha_\text{L}$ and $\alpha_\text{H}$, as in  Eq.~\eqref{eq:sf_eff_b}.
If we would assume a mass independence of the efficiency,
i.e. $\alpha_\text{L}=0$, for low mass
halos, the resulting star formation rate becomes much too large at
$z \gtrsim 3$ as seen in Fig.~\ref{fig:sfr}; without suppression
of the formation efficiency
too much more stars are produced in small halos.
On the other hand, for $\alpha_\text{H}=0$ the estimated star formation rate differs little from the default case.


\subsection{Reionization history}

\begin{figure}
   \centering
   \includegraphics[width=\columnwidth]{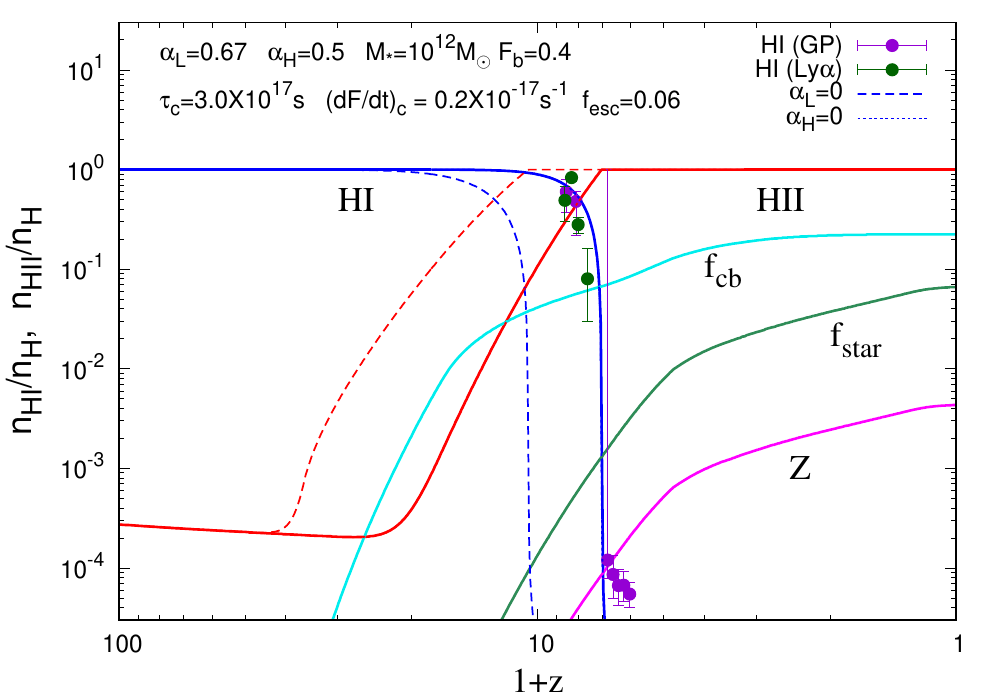}. 
   \caption{Fractions of HI(blue) and HII(red) against total hydrogen. 
     Filled circles show the volume-averaged neutral fraction from
     the Gunn-Peterson test~\protect\citep{Fan:2005es,Davies:2018yfp} (violet) and Ly$\alpha$ emission~\protect\citep{2020ApJ...904..144J,Morales:2021fzt} (green).
     The evolution of HI and HII for  $\alpha_\text{L}=0$ or $\alpha_\text{H}=0$
     is shown by dashed or dotted lines.
     The evolution of metallicity $Z$ (magenta),
     the fraction $f_\text{star}$ of baryons used for star formation (geen)
     and the fraction $f_\text{cb}$ of cooled baryons that could be used
     to form stars
     are also shown (light blue line).}
   \label{fig:ionization_H}  	
\end{figure}

\begin{figure}
   \centering
   \includegraphics[width=\columnwidth]{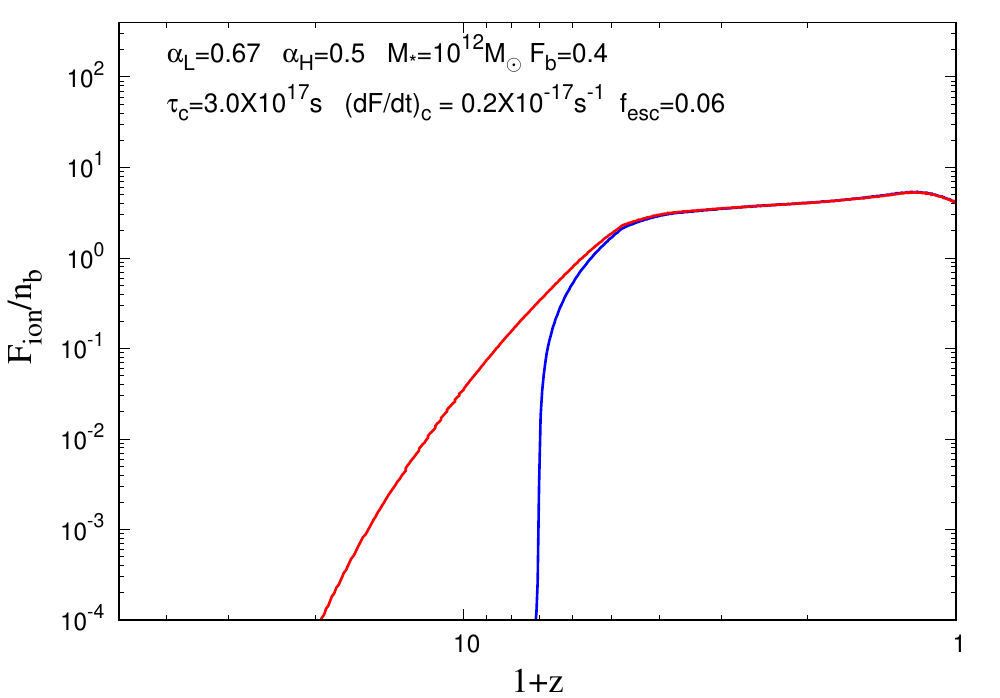}. 
   \caption{Ratio of the effective ionizing flux to the baryon number density.
     The red (blue) solid line represents the cumulative value without (with)
     absorption by the intergalactic medium. }
   \label{fig:ionizing_flux}  	
\end{figure}

The UV flux from stars~\eqref{eq:uv_flux} heats the background plasma and
ionizes 
hydrogen and helium in the intergalactic medium.
We calculate the reionization history and the evolution of
the electron temperature.

Fig.~\ref{fig:ionization_H} shows the reionization history.
The blue and red solid lines denote the fraction of HI and HII, respectively.
The observed HI fraction from measurements of the Gunn-Perterson
optical depth~\citep{Fan:2005es,Davies:2018yfp} and Lyman-alpha emission~\citep{2020ApJ...904..144J,Morales:2021fzt} is also shown. 
The model shows that the HI fraction ($\sim 70\%$)  at $z\sim 7$ is larger than the observation from the Lyman-alpha emission.
On the other hand, the recent observation of the mean free path of ionising photons~\citep{Becker:2021jyx} implies that the HI fraction is larger than 20\% at $z=6$.
The uncertainty is still significant in the observed HI fraction around these redshifts. 
We note that in a highly dense region UV flux
may be screened and HI remains to lower redshift. This is not taken into
account in our calculation.
We also show the evolution of the cumulative ionising
flux~\eqref{eq:ionization_flux}, emitted and after absorbed by IGM,
in Fig.~\ref{fig:ionizing_flux} below.

In Figs.~\ref{fig:ionization_H} we see that reionization takes place
around $z\simeq 6.5$ when the ionization flux reaches a fraction ($0.2-0.3$)
of the baryon number
density (Fig.~\ref{fig:ionizing_flux}).
The resulting reionization epoch agrees with the observed
optical depth in QSO lights~\citep{Fan:2005es,Davies:2018yfp}.

The reionization history for $\alpha_\text{L}=0$ or $\alpha_\text{H}=0$
is also depicted by dashed or dotted lines.
With  $\alpha_\text{H}=0$  
the electron temperature raised earlier as in Fig.~\ref{fig:jeans_mass},
the ionization fraction of the hydrogen increases earlier, and the
reionization takes place around $z\simeq 10$.
On the other hand, with $\alpha_\text{L}=0$ 
the increase of star formation efficiency
for large haloes hardly
affects the reionization history; the line is basically
degenerate with the default curve.

\begin{figure}
   \centering
   \includegraphics[width=\columnwidth]{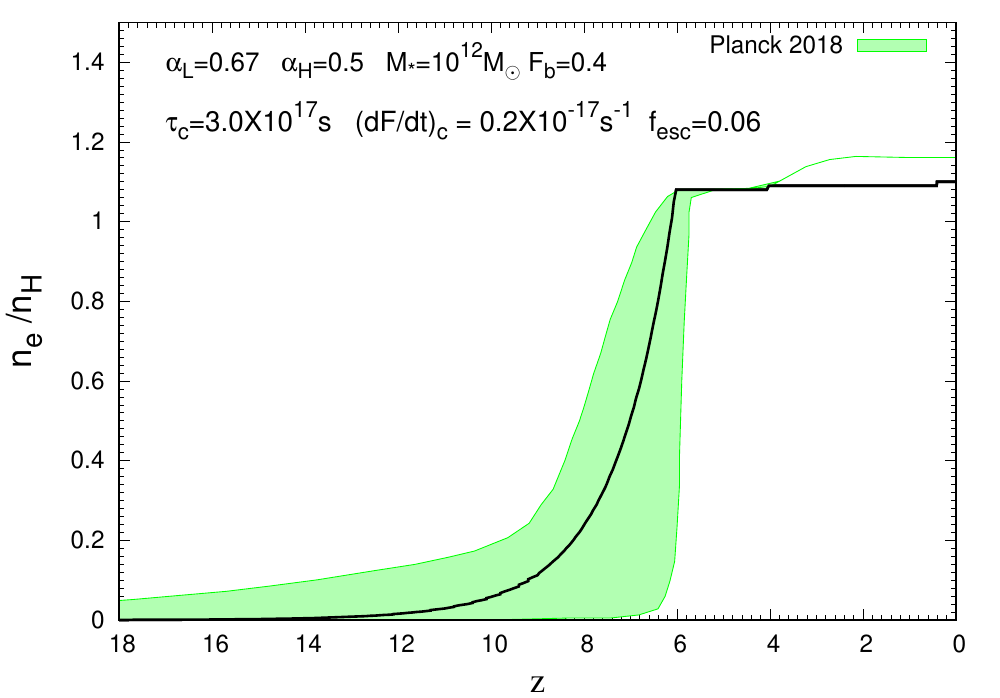}. 
   \caption{Evolution of the free electron fraction (black line)
     with respect to the hydrogen amount
     compared with the constraint from Plank
     (green region)~\protect\citep{Planck:2018vyg}. }
   \label{fig:free_el_frac}    
\end{figure}

With this reionization history we calculate the optical depth $\tau= 0.051$,
which agrees with the Planck 2018 value,
$\tau = 0.054\pm 0.007$~\citep{Planck:2018vyg}. 
Fig.\ref{fig:free_el_frac} shows the evolution of the free electron
fraction compared with that constrained from 
the Planck observation (Fig.45 of~\citet{Planck:2018vyg}).
The free electron fraction derived from 
the Planck data show some increase at $z=3.5$ towards
lower redshift.
This is because the Planck analysis assumes the full
ionization of helium at $z\lesssim 3.5$, which gives a negligible
contribution to $\tau$.
Taking this into account, the free electron fraction is
in a good agreement with the Plank observation.

In Fig.~\ref{fig:ionization_H} we also plot the evolution of
metallicity ($Z$), the fraction of baryons that have experienced star
formation ($f_\text{star}$), and the fraction of the cooled baryons
against the total ($f_\text{cb}$). We see that stars are 
about 1/4 the baryons that satisfy the cooling conditions.
Star formation starts around $z\gtrsim 10$.  A significant fraction
of stars ($\sim 40\%$) formed between $z=1$ and $z=3$. 
The final fraction of stars amounts to $f_\text{star} \simeq 0.07$,
or $\Omega_\text{star} \simeq 0.004$.
In~\citet{Fukugita:2004ee} it is estimated that fraction of baryons
that undergo stars is $\Omega_\text{star}=0.0027\times 1.62=0.0044$,
where the first number is the present day mass density of stars,
and the second is a correction for recycling.
The metallicity evolves in a similar way and the global mean of
present-day metallicity is $Z\simeq 0.005$.

\subsection{Luminosity function}

\begin{figure}
   \centering
   \includegraphics[width=\columnwidth]{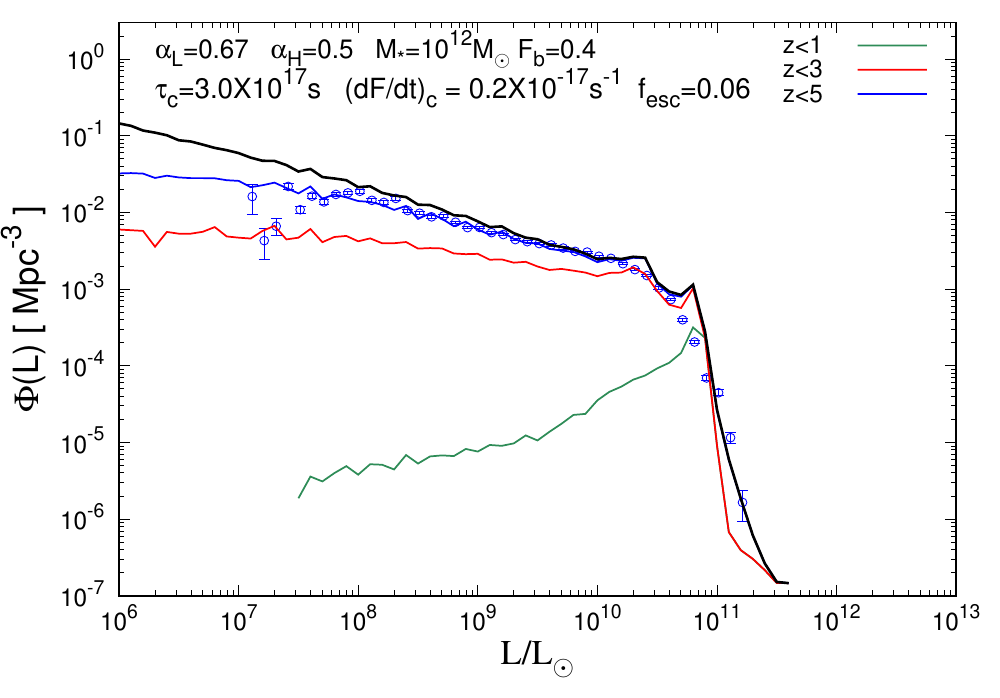}. 
   \caption{Present-day luminosity function of galaxies (black line).
     We indicate the luminosity functions of halos formed
     at $z<5$ (blue), $z<3$ (red) and $z<1$ (green). The data points are
   taken from  the SDSS observation of~\protect\cite{Blanton:2004zy}.}
   \label{fig:luminosity_func}      
\end{figure}

In Fig.~\ref{fig:luminosity_func} we show the present-day luminosity
function of galaxies with thick black line using
Eq.~\eqref{eq:luminosity_func}.
We show the SDSS observation~\citep{Blanton:2004zy} with data points.
How much haloes formed at $z<5$, $z<3$ and  $z<1$  contribute
to the present-day luminosity function are also indicated.
The derived luminosity function (black line) is broadly consistent
with the observation, including the overall normalization (up to a factor
of $<2$) and the turn off of the
luminosity function.
It is found that most contribution comes
from galaxies formed between $z=5$ and $z=1$.  A significant fraction
of low luminosity galaxies with $L \lesssim 10^9 L_\odot$ still form 
at $5>z > 3$.



\begin{figure}
   \centering
   \includegraphics[width=\columnwidth]{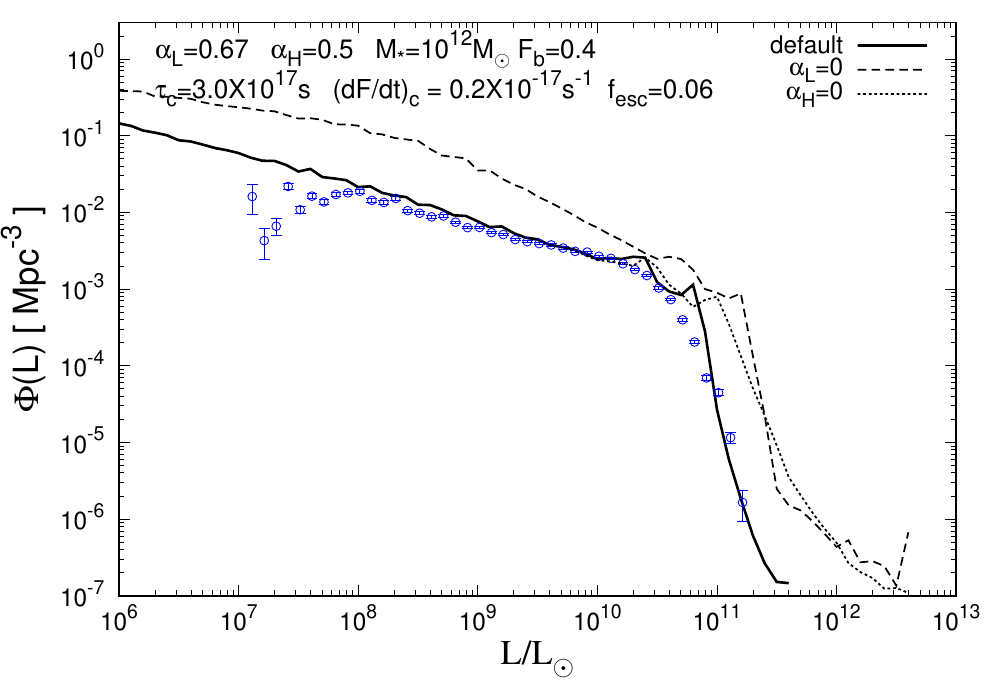}. 
   \caption{Present-day luminosity functions for
     $(\alpha_\text{L},\alpha_\text{H})=(0.7,1.4)$ (solid), $(0,1.4)$ (dashed)
     and  $(0.7,0)$ (dotted). }
   \label{fig:luminosity_func_cp}  	
\end{figure}

In Fig.~\ref{fig:luminosity_func_cp} we show, for comparison,
the luminosity function
for different star formation efficiencies, in addition to our default
choice $(\alpha_\text{L}, \alpha_\text{H}) = (0.67,0.5)$, the two cases with
more formation in earlier $(0,0.5)$ or later epoch $(0.67,0)$.
For $\alpha_\text{L}=0$ the luminosity function becomes too large,
especially at low luminosity. 
On the other hand, if we take $\alpha_\text{H}=0$, the model gives
too large a number of high luminosity galaxies, the turn-off luminosity
largely shifted to a brighter side. 
The agreement with the observation is achieved only by introducing  
suppression factors in the star formation rate in both high and low mass
haloes. 

The compatibility of both star formation rate and luminosity function
with the observation mean
that the star formation efficiency should have a peak around $M_*=10^{12}M_\odot$
and suppressed for both $M < M_*$ and $M > M_*$.
Our choice $\alpha_\text{L}=0.67$ and $M_*$ agree with~\citet{Behroozi:2012sp} who give $\alpha_\text{L}\sim 2/3$ and $\alpha_\text{H} \sim 4/3$ with $M_* \simeq 10^{11.7}M_\odot$.
On the other hand, our model shows a weaker suppression for $M>M_*$.
This is due to the fact that the ineffective metal cooling suppresses the star formation in addition to the factor of $(M/M_*)^{-\alpha_\text{H}}$, which makes our $\alpha_\text{H}$ smaller than that in~\citet{Behroozi:2012sp}.

\section{Conclusion}
\label{sec:conclusion}

We have studied the reionization history, star formation history, and
the present luminosity function of galaxies using the simplest model of 
galaxy formation in $\Lambda$CDM cosmology. 
We have found that the simple semi-analytic model,
a la Press and Schecheter (1974), can account for
the observed reionization epoch, the star formation rate and
the present-day luminosity function to almost a quantitative accuracy, if the physical processes,
the Jeans condition and the cooling conditions are properly
taken into account.
The total amount of stars formed are estimated to be $\Omega_\mathrm{star}=0.004$ compared to the observation $0.0044$ including the recycling factor.
The epoch of reionisation 
is etimated to occur sharply at $1+z\simeq 7.5$.

To account for the observed time dependence of the star formation
rate and for the present luminosity function of galaxies,
both independently requires that
the star formation efficiency should 
increase with the halo mass for $M<M_*$, it reaches a maximum
at the turn-off mass of $M_* \sim 10^{12}M_\odot$, and then it should
decrease towards
larger halo mass. Namely, there is an optimal mass for star formation,
$M\sim 10^{12}M_\odot$ for the halo mass, or $M\sim 10^{10}M_\odot$ for the stellar
mass of galaxies. This is the only constraints to be added to the model
of the collapse.

Our simple model accounts for the observed star formation
history and the present-day luminosity function by choosing  appropriately a set of $7$ parameters.
Among them $f_\text{esc}$, $F_b$ and $\dot{F}_\text{c}\tau_{c}$ are taken consistently with the direct observations of the escape fraction and the age distribution of the stars in the solar neighborhood.
The three more parameters that control the mass-dependent star formation
rate, $\alpha_H$, $\alpha_L$, and $M_*$, are determined to make the result
consistent with the observation.


\section*{Acknowledgements}
This work was supported by JSPS KAKENHI Grant Nos. 20H05851(M.K.), 21K03567(M.K.), and World Premier International Research Center Initiative (WPI Initiative), MEXT, Japan (M.F, M.K.).

\section*{Data availability}
No new data were generated or analysed in this article. 
The code that we developed to produce the results presented in this manuscript is undergoing further extension, and we plan to publicly release it at a later stage as part of a follow-up work.

\bibliographystyle{mnras}
\bibliography{reionize}

\end{document}